\documentstyle[12pt]{article}
\def\beq{\begin{equation}}
\def\eeq{\end{equation}}
\def\bea{\begin{eqnarray}}
\def\eea{\end{eqnarray}}
\def\bq{\begin{quote}}
\def\eq{\end{quote}}

\def\gappeq{\mathrel{\rlap {\raise.5ex\hbox{$>$}}
{\lower.5ex\hbox{$\sim$}}}}

\def\lappeq{\mathrel{\rlap{\raise.5ex\hbox{$<$}}
{\lower.5ex\hbox{$\sim$}}}}

\begin{document}
\topmargin -0.5cm
\oddsidemargin -0.8cm
\evensidemargin -0.8cm
\pagestyle{empty}
\begin{flushright}
{CERN-TH.7013/93}
\end{flushright}
\vspace*{5mm}
\begin{center}
{\bf NON-PERTURBATIVE EFFECTIVE LAGRANGIANS FOR} \\
{\bf SUPER-MATRIX MODELS } \\
\vspace*{0.3cm}
{\bf Ram Brustein}$^{*)}$ \\
\vspace{0.3cm}
Theoretical Physics Division, CERN \\
CH - 1211 Geneva 23\\
and\\
Dept. of Physics, Univ. of Pennsylvania\\
Philadelphia, PA 19104, U.S.A.\\
\vspace*{0.3cm}
{\bf Michael Faux}$^{*)}$ and {\bf Burt A. Ovrut}$^{**),***)}$ \\
\vspace{0.3cm}
Theoretical Physics Division, CERN \\
CH - 1211 Geneva 23\\
\vspace*{2cm}
{\bf ABSTRACT}
\end{center}
\vspace*{3mm}
\noindent
We discuss $d=1, {\cal N}=2$ supersymmetric matrix models
and exhibit the associated $d=2$ collective field theory in the
limit of dense eigenvalues.  From this field theory we construct,
by the addition of several new fields, a $d=2$ supersymmetric
effective field theory, which reduces to the collective field
theory when the new fields are replaced with their vacuum
expectation values.  This effective theory is Poincare invariant
and contains perturbative and non-perturbative information
about the associated superstrings.

\vspace*{1.8cm}
\noindent
\rule[.1in]{16.5cm}{.002in}\\
\noindent
$^{*)}$ Work supported in part by DOE under Contract No.
DOE-AC02-76-ERO-3071.\\
$^{**)}$ On sabbatical leave from the Department of Physics, University
of Pennsylvania,
Philadelphia, PA 19104, U.S.A.\\
$^{***)}$ Invited talk, International Europhysics Conference on High
Energy Physics, Marseille,
France, July 22-28, 1993.
 \vspace*{0.5cm}

\begin{flushleft} CERN-TH.7013/93 \\
September 1993
\end{flushleft}
\vfill\eject

\setcounter{page}{1}
\pagestyle{plain}

\def\baselinestretch{1.3}
\setlength{\oddsidemargin}{-1.0cm}
\setlength{\textwidth}{17.0cm}
\setlength{\topmargin}{-.7cm}
\setlength{\textheight}{24.0cm}%
\renewcommand{\theequation}{\arabic{equation}}
\def\bl{\Biggl\{}
\def\br{\Biggr\}}
\def\fl{\flushleft}
\def\fr{\flushright}
\def\L{{\cal L}}
\def\R{\cal{R}}
\def\a{\alpha}
\def\b{\beta}
\def\d{\delta}
\def\e{\epsilon}
\def\G{\Gamma}
\def\g{\gamma}
\def\l{\lambda}
\def\n{\eta}
\def\r{\rho}
\def\z{\zeta}
\def\tPhi{\tilde{\Phi}}
\def\O{\cal{O}}
\def\T{\theta}
\def\s{\sigma}
\def\k{\kappa}
\def\t{\theta}
\def\vphi{\varphi}
\def\w{\omega}
\def\bQ{\bar{Q}}
\def\be{\bar{\epsilon}}
\def\bn{\bar{\eta}}
\def\bpsi{\bar{\psi}}
\def\bT{\bar{\T}}
\def\bD{\bar{D}}
\def\hf{\frac{1}{2}}
\def\der{\partial}
\def\bq{\begin{equation}}
\def\eq{\end{equation}}
\def\brr{\begin{eqnarray}}
\def\err{\end{eqnarray}}
\def\ba{\left(\begin{array}}
\def\ea{\end{array}\right)}
\def\pp{\hbox{\ooalign{$\displaystyle\int$\cr$-$}}}
\def\derbar{\stackrel{\leftrightarrow}{\partial}}
\def\dd{\stackrel{\leftrightarrow}{\partial}}
\def\ba{\left(\begin{array}}
\def\ea{\end{array}\right)}
\def\gbig {\hbox{\Large\it g}}

The question of how supersymmetry is broken remains one of the most
important issues in superstring theory.  The advent of $d=1$
supersymmetric matrix models\cite{marpar},
which contain nonperturbative
information about a class of two-dimensional superstrings, offers the
hope that one can identify, and study quantitatively, the exact
string mechanism for supersymmetry breaking.  To do this, it is
essential
to construct the complete two-dimensional
supersymmetric effective Lagrangian associated with this class
of theories.  Once that is achieved, one can look for non-perturbative
phenomena, such as instantons,
that will induce supersymmetry violating operators (spontaneous or
explicit) in this effective Lagrangian.  This entire program has
already been carried out\cite{bo}, with positive results, in the case of
$d=1$ bosonic matrix models where the nonperturbative effective
Lagrangian was constructed and its fundamental
symmetry, a non-compact shift symmetry, $\z\rightarrow\z+c$
in one of its fields $\z$, was shown to be broken by
instantons in a single eigenvalue of the matrix model.  In this talk
we will discuss the extension of these results to the supersymmetric
case\cite{bfo}.  To be precise, we will construct the
two-dimensional supersymmetric effective Lagrangian for the
superstrings associated with the $d=1$ supersymmetric matrix model.  We
have also discovered instantons in the eigenvalues of these matrix
models, which have all the right characteristics to break supersymmetry.
However, the constraints placed on the length of this manuscript
mandate that we discuss these instantons and supersymmetry breaking
elsewhere. \\
\indent The simplest supersymmetric matrix model which admits
non-trivial
interactions involves a $d=1, {\cal N}=2$ supersymmetry.  The
fundamental variable of the theory is a time-dependent
$N\times N$, Hermitian matrix superfield
\bq \Phi_{ij}\!=\!M_{ij}(t)\!+\!i\t_1\Psi_{1ij}(t)\!+\!i\t_2\Psi_{2ij}
    +i\t_1\t_2F_{ij}(t),
 \eq
where $\t_1$ and $\t_2$ are the two anticommuting parameters.
The Lagrangian is given by
\bq L=\int d\t_1d\t_2\bl\hf Tr D_1\Phi D_2\Phi+iW(\Phi)\br,
 \eq
where the superpotential is
$W(\Phi)=\sum_n b_n Tr\Phi^n$.
Eliminating the auxiliary fields $F_{ij}$, quantizing the theory
and restricting the Hilbert space to the $U(N)$ invariant states,
one arrives\cite{dabh,bfo} at an effective component field Lagrangian of
the
 form
\brr L &=& \sum_i\{\hf\dot{\l}_i^2
     -\hf(\frac{\der W_{eff}(\l)}{\der\l_i})^2\}
     \nonumber \\
     & &\hspace{-.4in} -\frac{i}{2}\sum_i(\psi_{1i}\dot{\psi}_{1i}
     +\psi_{2i}\dot{\psi}_{2i})
-i\sum_{ij}\psi_{1i}\frac{\der^2W_{eff}(\l)}{\der\l_i\der\l_j}
     \psi_{2j}.\hspace{.2in}
 \label{lag} \err
where $W_{eff}=W+w$ and
$w(\l)=-\sum_{i}\sum_{j\ne i}\ln|\l_i-\l_j|$.
The $\l_i$ are the $N$ eigenvalues of $M_{ij}$ and $\psi_i$
are associated fermions.
Lagrangian (\ref{lag}) is $d=1, {\cal N}=2$
supersymmetric.  This theory can be converted from one dimensional
quantum mechanics to a two-dimensional field theory
by introducing a real parameter, $x$, restricted to the interval
$-L/2<x<L/2$ and by defining the collective fields
\brr \vphi(x,t) &=& \sum_i\Theta(x-\l_i(t)) \nonumber \\
     \psi(x,t) &=& -\sum_i\d(x-\l_i(t))\chi_{i}(t).
 \label{colldef} \err
The limit of physical interest is when $N\rightarrow\infty,
L\rightarrow\infty$. See\cite{prev} for previous work on this subject.
If $N/L\rightarrow\infty$, then the
eigenvalues become dense in $L$ and the collective fields
(\ref{colldef}) become unconstrained space-time fields.  In this
case, rewriting (\ref{lag}) in terms of the collective fields
and appropriately taking the $N\rightarrow\infty, L\rightarrow\infty$
limit, one finds that the Lagrangian becomes
\brr L &=& \int dx\bl\frac{\dot{\vphi}^2}{2\vphi'}
     -\frac{\pi^2}{6}\vphi^{'3}
     +\hf\w^2x^2\vphi' \nonumber \\
     & &
     -\frac{i}{2\vphi'}(\psi_1\dot{\psi}_1+\psi_2\dot{\psi}_2)
     -\frac{i\pi}{2}\psi_1\psi_1'
     +\frac{i\pi}{2}\psi_2\psi_2' \nonumber \\
     & &
     +\frac{i}{2}\frac{\dot{\vphi}}{\vphi^{'2}}
     (\psi_1\psi_1'+\psi_2\psi_2')\br,
 \label{lagcollnoncan} \err
where $\psi=\frac{1}{\sqrt{2}}(\psi_1+i\psi_2)$
and $\w^2$ is an arbitrary positive constant.
The equations of motion associated with (\ref{lagcollnoncan})
have a static solution with $<\psi_1>=<\psi_2>=0$ and
$<\vphi'>=\frac{1}{\pi}\sqrt{\w^2x^2-\frac{1}{{\rm g}}}$,
where ${\rm g}$ is an arbitrary positive integration constant.
In this case, $<\vphi>$ is only defined in the region
$|x|\ge 1/\w\sqrt{{\rm g}}$.  Therefore, in this region only,
we can expand $\vphi$ as $\vphi=<\vphi>+\frac{1}{\sqrt{\pi}}\z$.
Furthermore, if we define a new spatial coordinate $\tau$
through the expression $\tau'=\frac{1}{\pi}<\vphi'>^{-1}$,
the Lagrangian (\ref{lagcollnoncan}) can be reexpressed as
\brr \lefteqn{L = \int d\tau\bl
       \hf(\dot{\z}^2-\z^{'2})} \nonumber \\
       & &-\hf\frac{\gbig(\tau)\dot{\z}^2\z'}{1+\gbig(\tau)\z'}
       -\frac{1}{6}\gbig(\tau)\z^{'3}
       +\frac{1}{3\gbig(\tau)^2} \nonumber \\
       & & -\frac{i}{\sqrt{2}}(\psi_+\dot{\psi}_+-\psi_+\psi_+')
       -\frac{i}{\sqrt{2}}(\psi_-\dot{\psi}_-+\psi_-\psi_-')
       \nonumber \\
       & &
       +\frac{i}{\sqrt{2}}\frac{\gbig(\tau)\z'}{1+\gbig(\tau)\z'}
       (\psi_+\dot{\psi}_++\psi_-\dot{\psi}_-) \nonumber \\
       & & +\frac{i}{\sqrt{2}}\frac{\gbig(\tau)\dot{\z}}
       {(1+\gbig(\tau)\z')^2}(\psi_+\psi_+'+\psi_-\psi_-')\br,
 \label{lagcollfinal} \err
where $\psi_{+,-}=\frac{2^{1/4}}{\sqrt{\pi}}\psi_{1,2}$,
$\gbig(\tau)$ is a space dependant coupling parameter given by
\bq \gbig(\tau)=\frac{ 4\sqrt{\pi}e^{-2\w(\tau-\tau_0)}         }
                 { \w(1-\frac{1}{\w{\rm g}}e^{-2\w(\tau-\tau_0)})^2 }.
 \label{gdef} \eq
and $\tau_0$ is an arbitrary constant.  Note that the Lagrangian
(\ref{lagcollfinal}) suffers from the fact that the coupling
parameter $\gbig(\tau)$ is explicitly a function of $\tau$ and, hence,
$L$ is not Poincare invariant.  To overcome this difficulty, we
make the assumption that $\gbig(\tau)$ arises as the vacuum expectation
value (VEV) of a function of a new scalar field, which we denote by
$\a$.
We infer the existence of an effective Poincare invariant theory
involving $\a$, as well as $\z, \psi_+$ and $\psi_-$, which
reproduces (\ref{lagcollfinal}) when $\a$ is replaced by its
$\tau$-dependant VEV.   Additionally, we postulate that this
effective field theory posesses a $d=2$ supersymmetry.  It follows
that, in addition to $\a$, we must introduce its fermionic
superpartners $\chi_+$ and $\chi_-$ which, of course, have
vanishing VEV'S.  There are an infinite
number of $d=2, (p,q)$ supersymmetries. It is not hard to
demonstrate that the appropriate supersymmetry for our effective
field theory is (1,1) supersymmetry.  The original and newly introduced
fields discussed above appear as components of the (1,1) superfields
\brr \Phi_1 &=& \z+i\t^+\psi_++i\t^-\psi_-+i\t^+\t^-Z  \nonumber \\
     \Phi_2 &=& \a+i\t^+\chi_++i\t^-\chi_-+i\t^+\t^-A.
 \err
The construction of the correct supersymmetric
effective field theory proceeds as follows. Consider the
free part of the collective field Lagrangian.
It is given by
\bq \L_{01}=\hf(\dot{\z}^2-\z^{'2})-i\psi_+\der_-\psi_+
         -i\psi_-\der_+\psi_-.
 \label{oops} \eq
Now consider the manifestly supersymmetric Lagrangian
\brr \lefteqn{\L_{01}^{(eff)}=
        \int d\t^+d\t^-D_+\Phi_1D_-\Phi_1} \nonumber \\
        \!\!\!\!\!\!\!\!&=&\!\!\!\!\!
        \hf(\dot{\z}^2\!-\!\z^{'2})\!-\!i\psi_+\der_-\psi_+
         \!-\!i\psi_-\der_+\psi_-\!+\!Z^2\!\!.
 \label{lago1} \err
Field $Z$ is auxiliary and can be eliminated using its equation of
motion, $<Z>=0$.  It is clear that
(\ref{lago1}) reproduces
(\ref{oops}).  We also need to introduce a kinetic
energy for $\Phi_2$.  To do this, we must find a superfield Lagrangian
involving $\Phi_2$ only, whose component field equations
admit $<\a>=\exp{(-\w(\tau-\tau_0))}, <\chi_\pm>=<A>=0$
as a solution, and which reduces to the $1/3\gbig(\tau)^2$ term
in (\ref{lagcollfinal}) when these VEV's are inserted.  The appropriate
Lagrangian is
\brr \L_{02}^{(eff)} \!\!\!&=&\!\!\! \int d\t^+d\t^-\bl
    F_1(\Phi_2)D_+\Phi_2D_-\Phi_2 \nonumber \\
    & &\!\!\!
    -\frac{1}{\w^2}F_2(\Phi_2)\der_-D_+\Phi_2\der_+D_-\Phi_2\br
 \label{lago2} \err
where $F_1$ and $F_2$ are polynomial functions
of $\Phi_2$ whose exact form is irrelevant for this discussion.
We proceed to study the interaction terms in  (\ref{lagcollfinal}).
The terms in
(\ref{lagcollfinal}) linear in the coupling  $\gbig(\tau)$ are
\brr \L_1 &=& \gbig(\tau)\bl-\frac{1}{6}(\z^{'3}+3\dot{\z}^2\z')
\nonumber \\
     & &\hspace{-.5in}
     +\frac{i}{\sqrt{2}}\z'(\psi_+\dot{\psi}_++\psi_-\dot{\psi}_-)
+\frac{i}{\sqrt{2}}\dot{\z}(\psi_+\psi_+'+\psi_-\psi_-')\br.\hspace{.3in}
 \label{laglin} \err
We would like to find a superfield Lagrangian involving
$\Phi_1$ and $\Phi_2$ which, when added to (\ref{lago1})
and (\ref{lago2}), continues to admit $<\a>=\exp{(-\w(\tau-\tau_0))}$,
$<\chi_\pm>=<A>=0$ and $<Z>=0$ as a solution to the equations of motion
and which, when these are substituted, reduces to
(\ref{laglin}).  We have shown that
the appropriate Lagrangian is \nopagebreak
\brr \lefteqn{ \!\!\!\!\!\L_1^{(eff)}\!\!\!=\!\!\!\int d\t^+d\t^-\bl}
     \nonumber \\
     \!\!\!\!\!& &\!\!\!
     \frac{f(\Phi_2)}{\w^3\Phi_2^3}
     \der_{(+}\Phi_1\der_{-)}\Phi_2\der_{[+}\Phi_1\der_{-]}\Phi_2
     D_{(+}\Phi_1D_{-)}\Phi_2
         \nonumber \\
     \!\!\!\!\!& &\!\!\!
     -\frac{f(\Phi_2)}{\w^5\Phi_2^5}
     (\der_{[+}\Phi_1\der_{-]}\Phi_2)^2\der_{(+}\Phi_1\der_{-)}\Phi_2
     D_+\Phi_2D_-\Phi_2
  \nonumber \\
     \!\!\!\!\!& &\!\!\!
     +\frac{1}{3}\frac{f(\Phi_2)}{\w^5\Phi_2^5}
     (\der_{(+}\Phi_1\der_{-)}\Phi_2)^3
     D_+\Phi_2D_-\Phi_2\br,
\err
where
\bq f(\Phi_2)=\frac{4\sqrt{\pi}}{\w}
    \frac{\Phi_2^2}{(1-\frac{1}{{\rm g}\w}\Phi_2^2)^2},
 \eq
$A_{(+}B_{-)}=A_+B_-+A_-B_+$, and $A_{[+}B_{-]}=A_+B_--A_-B_+$.
It is possible to extend this expression to
all orders in
$\gbig(\tau)$,
\brr \lefteqn{\L^{(eff)} = \int d\t_+d\t_-\bl
     D_+\Phi_1D_-\Phi_1 } \nonumber \\
     & &\hspace{-.2in}
    +F_1(\Phi_2)D_+\Phi_2D_-\Phi_2
    \!-\!\frac{1}{\w^2}F_2(\Phi_2)
    \der_-D_+\Phi_2\der_+D_-\Phi_2 \nonumber \\
     & &\hspace{-.2in}
     +\frac{f(\Phi_2)}{\w^3\Phi_2^3}
     \frac{\der_{(+}\Phi_1\der_{-)}\Phi_2
           \der_{[+}\Phi_1\der_{-]}\Phi_2}
          {1+\frac{f(\Phi_2)}{\w\Phi_2}
           \der_{(+}\Phi_1\der_{-)}\Phi_2}
           D_{(+}\Phi_1D_{-)}\Phi_2  \nonumber \\
     & &\hspace{-.3in}
     -\frac{f(\Phi_2)}{\w^5\Phi_2^5}
     \frac{(\der_{[+}\Phi_1\der_{-]}\Phi_2)^2
     \der_{(+}\Phi_1\der_{-)}\Phi_2}
     {1+\frac{f(\Phi_2)}{\w\Phi_2}
     \der_{(+}\Phi_1\der_{-)}\Phi_2}
     D_+\Phi_2D_-\Phi_2
\nonumber \\
     & &\hspace{-.2in}
     +\frac{1}{3}\frac{f(\Phi_2)}{\w^5\Phi_2^5}
     (\der_{[+}\Phi_1\der_{-]}\Phi_2)^3
     D_+\Phi_2D_-\Phi_2\br. \label{last} \err
The component field equations of motion
have $<\a>=\exp{(-\w(\tau-\tau_0))},\ \ <\chi_\pm>=<A>=0$
and $<Z>=0$ as a solution.  When these are substituted back
into the component field expansion of (\ref{last}), one obtains
exactly the collective field Lagrangian, (\ref{lagcollfinal}).
Therefore, the Lagrangian (\ref{last}) is the
two-dimensional supersymmetric effective Lagrangian for the superstrings
associated with the $d=1$ supersymmetric matrix model.

\end{document}